\documentclass[aps,pra,twocolumn,groupedaddress,showpacs]{revtex4}
\usepackage{graphicx}
\usepackage[ansinew]{inputenc}
\usepackage{array}
\usepackage{color}
\usepackage{amsmath}
\usepackage{amsxtra}
\usepackage{amssymb}
\usepackage{latexsym}
\usepackage{dsfont}
\begin{document}
\title{Optomechanical trapping and cooling of partially transparent
mirrors}
\author{M. Bhattacharya, H. Uys, and P. Meystre}
\affiliation{B2 Institute, Department of Physics and College of
Optical Sciences, The University of Arizona, Tucson, Arizona
85721}

\date{\today}

\begin{abstract}
We consider the radiative trapping and cooling of a partially
transmitting mirror suspended inside an optical cavity,
generalizing the case of a perfectly reflecting mirror previously
considered [M. Bhattacharya and P. Meystre, Phys. Rev. Lett.
\textbf{99}, 073601 (2007)]. This configuration was recently used
in an experiment to cool a nanometers-thick membrane [Thompson
\textit{et al.}, arXiv:0707.1724v2, 2007]. The self-consistent
cavity field modes of this system depend strongly on the position
of the middle mirror, leading to important qualitative differences
in the radiation pressure effects: in one case, the situation is
similar that of a perfectly reflecting middle mirror, with only
minor quantitative modifications. In addition, we also identify a
range of mirror positions for which the radiation-mirror coupling
becomes purely dispersive and the back-action effects that usually
lead to cooling are absent, although the mirror can still be
optically trapped. The existence of these two regimes leads us to
propose a bichromatic scheme that optimizes the cooling and
trapping of partially transmissive mirrors.
\end{abstract}


\maketitle
\section{Introduction}

\begin{figure*}
\includegraphics[width=0.99 \textwidth]{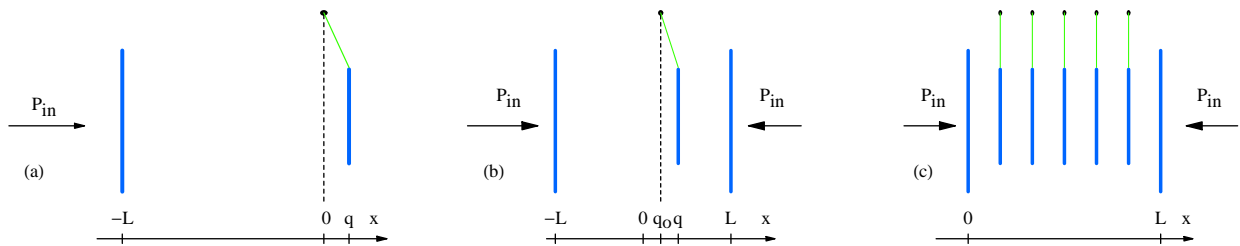}
\caption{\label{fig:transpic}(Color online).(a) The typical layout for
optomechanical cooling and trapping using a two-mirror cavity (2MC).
(b) A layout recently suggested by the authors for the same purpose
using a three-mirror cavity (3MC) \cite{mishpm1} with a perfectly
reflecting middle mirror, and implemented experimentally in
\cite{thompson2007} using a partially transparent dielectric membrane
in place of the middle mirror. (c) A possible arrangement for
scaling the technique to more than one mirror. The parameters
labelling the figures are defined in the text.}
\end{figure*}

The optomechanical cooling and trapping of mirrors has recently
become the subject of an intense research effort as it promises to
offer a viable means of extending quantum mechanics to macroscopic
objects
\cite{gigan2006,kleckner2006,arcizet2006,schliesser2006,corbitt2007}.
The typical experimental arrangement consists of a linear
two-mirror optical cavity (2MC) driven by laser radiation close to
a cavity resonance [Fig.\ref{fig:transpic}(a)]. One of the mirrors
in the cavity is small and is mounted on a cantilever, so as to
be movable, and the goal is to cool its vibrational state of
motion to a point as close to its quantum mechanical ground state
as possible. The cooling proceeds with the use of two laser beams,
the first one detuned to the blue of a cavity resonance and
providing an optical trap for the movable mirror, with a frequency
$\omega_{\rm eff}$ larger than the intrinsic cantilever frequency
$\omega_M$; and the second one detuned to the red of the cavity,
so as to (almost) independently increase the damping constant of
the oscillating mirror from its field-free value $D_M$ to
$D_{\rm eff}$ \cite{corbitt2007}.

From the quantum mechanical point of view, the combined effect of
the laser fields on the moving mirror is two-fold: they create
a harmonic trap with large energy level spacing $\hbar \omega_{\rm
eff}$, and cool the mirror from its initial equilibrium
temperature $T_{\rm e}$ to a lower value
\begin{equation}
\label{eq:lowT} T_{\rm eff}=\left(\frac{D_M}{D_{\rm eff}}\right)T_{\rm e}.
\end{equation}
as shown explicitly in Appendix \ref{sec:appa}. The trapping and
cooling effects thus lower the number of quanta of vibrational
excitation of the oscillating mirror to
\begin{equation}
\label{eq:nosc} n_M=\frac{k_{\rm B} T_{\rm eff}}{\hbar \omega_{\rm
eff}} =\frac{k_{\rm B} T_{\rm e}}{\hbar \omega_{\rm eff}}
\left(\frac{D_M}{D_{\rm eff}}\right),
\end{equation}
where $k_{\rm B}$ is Boltzmann's constant. Current experimental
effort is intensely focused at achieving $n_M < 1$, i.e. at
placing the mirror in its quantum mechanical ground state. We note
at the outset that the expression for $n_M$ given in
Ref.~\cite{mishpm1} includes an additional term
$(\Omega_M/\Omega_{\rm eff})^3$ as compared to
Eq.~(\ref{eq:nosc}). That additional term results from expressing
$n_M$ in terms of the `bare' oscillation frequency of the moving
mirror rather than its effective frequency and underestimates its
degree of excitation. The correct formula is Eq.~(\ref{eq:nosc})
of this article. However the correction does not bring any 
qualitative change to the conclusions of the earlier work, 
Ref.\cite{mishpm1}. The same correction has recently been realized by
other authors \cite{karrai2004,marquardt2007}.

In the 2MC the allowed laser power is limited by the onset of
mirror bistability \cite{dorsel1983}, placing bounds on the
achievable cooling and trapping. In addition, in that geometry
radiation pressure is not used optimally as it couples to the
mirror from one side only. Most importantly perhaps, the 2MC
requires the movable mirror to be one of the end-mirrors of a high
finesse cavity and to have a high mechanical quality as well.
Technically these are conflicting demands because the high finesse
that maximizes the cooling effect of radiation is best achievable
with massive, rigidly fixed mirrors. On the other hand, the high
mechanical quality that minimizes the oscillator's coupling to
thermal noise is best achievable with small, flexibly mounted
mirrors. These opposing requirements represent the main
experimental challenge to achieving states of vibration of low
quantum number in the 2MC.

In a recent article we proposed an alternative geometry that
allows one to reach and detect lower $n_M$'s for comparable
parameters \cite{mishpm1} by suspending a perfectly reflecting
mirror in the middle of a two-mirror cavity
[Fig.\ref{fig:transpic}(b)]. This three-mirror cavity (3MC)
arrangement was shown to possess at least three advantages over
the 2MC. First, it provides a higher value of $\omega_{\rm eff}$
for the mirror \cite{pm1985,khalili2001,vogel2003,sheard2004},
leading to fewer quanta of excitation, see Eq.~(\ref{eq:nosc}).
Second it removes bistability problems completely as far as the
trapping fields are concerned, and partially for the cooling
fields. Lastly, it increases the time available for observing the
quantum dynamics of the mirror before the onset of thermal
decoherence.

The present paper generalizes the analysis of the 3MC to the case
where the middle mirror is partially transmitting, with the goal
of determining to what extent its advantages are retained in that
case, and also as a first step toward determining whether the same
linear cavity can be used to quantize the motion of more than one
mirror, see [Fig.\ref{fig:transpic}(c)]. A classical treatment of
the 3MC was presented earlier \cite{pm1985}, however the noise
analysis did not include the vacuum fluctuations in the laser
fields and $\omega_{\rm eff}$ and $D_{\rm eff}$ were derived only
in the static (zero-frequency) limit. Since we are concerned with
cooling the movable mirror to its quantum mechanical ground state,
a full quantized treatment is clearly needed. We derive
expressions for the effective frequency and damping constant valid
for any frequency $\omega$, and in making contact with the case of
the perfectly reflecting middle mirror  we include details that
could not be presented in Ref.~\cite{mishpm1} for lack of space.

In the course of concluding this work we became aware of a recent
experiment that beautifully demonstrates the working of the 3MC
and points out some of its additional virtues \cite{thompson2007}.
In that work Thompson \textit{et al.} cooled a 50nm thick
dielectric membrane placed inside an optical cavity from room
temperature (294K) down to 6.82mK, i.e. by a factor of $4.4\times
10^{4}$. These authors also pointed out that the 3MC solves a
number of the technological challenges faced by the 2MC as it
allocates the requirements of high optical finesse and high
mechanical quality to different parts of the cavity. The high
finesse optical cavity now consists of two rigidly fixed mirrors,
while the suspended middle mirror (or membrane) can independently
have a high mechanical quality.  Additionally the 2MC only allows
the measurement of the mirror displacement $q$, while the 3MC
allows the measurement of $q^{2}$, thereby projecting the state of
the mirror into an energy eigenstate \cite{braginsky1980}.

Rather than elaborating on the salient features of
Ref.~\cite{thompson2007}, this paper examines the effect of
middle-mirror transparency on the bistability, effective trapping
frequency and damping displayed by the 3MC. Our model is
comprehensive in that it is valid for a totally as well as for a
partially reflecting mirror and also for an arbitrary placement of
the moving mirror inside the cavity. The consideration of various
limiting cases allows us to propose a new two-color scheme that
optimizes the cooling and trapping of the transparent mirror. We note
that issues similar to those considered in this work have recently
been presented in Ref.~\cite{karrai2007} using a different
formalism.

The paper is organized as follows. Section~\ref{sec:Ham} derives a
Hamiltonian of the moving mirror-cavity system valid in situations
where it is sufficient to consider two modes of the cavity field,
and Section~\ref{sec:PRL} shows how that Hamiltonian reduces to
the case of a perfectly reflecting middle mirror \cite{mishpm1}.
We then turn to the case of a finite transmission, with
Section~\ref{sec:Tnotzero} discussing the situation when the
moving mirror location yields a linear coupling to the photon
number difference in the two field modes, and
Section~\ref{sec:extrememirr} to the case where that coupling
becomes quadratic. Section~\ref{eff-freq} discusses the
modification of the oscillation frequency and damping rate of the
mirror by radiation pressure, Section~\ref{sec:propbi} applies
these results to the formulation of a proposal for a new trapping
and cooling configuration, and Section~\ref{sec:conclusion} is a
summary and conclusion. Appendix \ref{sec:appa} contains a careful
derivation of Eqs.(\ref{eq:lowT}) and (\ref{eq:nosc}), and
appendices \ref{sec:appb} and \ref{sec:appc} present details of
the cases considered in Sections~\ref{sec:Tnotzero} and
~\ref{sec:extrememirr}.

\section{Model Hamiltonian}
\label{sec:Ham} We consider a 3MC geometry with the outer mirrors
fixed at $x=\pm L$ [Fig.\ref{fig:transpic}(b)] and a middle mirror
of transmissivity $T$ located at a position $x=q$. We assume the
mirror thickness to be much smaller than an optical wavelength,
a condition that has been realized experimentally \cite{thompson2007}.

\subsection{Classical modes}
\label{subsec:class}

We proceed by first determining the mode frequencies of the full
resonator as a function of $T$ and $q$. In the simple case $T=0,
q=0$ the resonant frequencies of the two sub-cavities are
\begin{equation}
\label{eq:wn}
\omega_{n}=\frac{n \pi c}{L},
\end{equation}
where
\begin{equation}
\label{eq:modeno} n= 2L/\lambda_n,
\end{equation}
$\lambda_n = 2\pi c/\omega_n$ and $n $ is the mode number (Table
\ref{tab:mirrorparam}).

When $T \neq 0$, the two sides of the resonator are coupled and
the modes of the complete system are found by solving the
Helmholtz equation with the appropriate boundary conditions at
$x=q,\pm L$, as described in Ref.~\cite{fader1985}. For this
calculation we assume for simplicity that the mirrors at $x=\pm L$
are perfectly reflecting. We also consider high-order cavity modes
such that $L \gg \lambda_n$ and mirror displacements $q$ (modulo
$\lambda_{n}) \ll \lambda_n$. The finite transmission of the
end-mirrors will be accounted for later on.

\begin{figure}
\includegraphics[width=0.5 \textwidth]{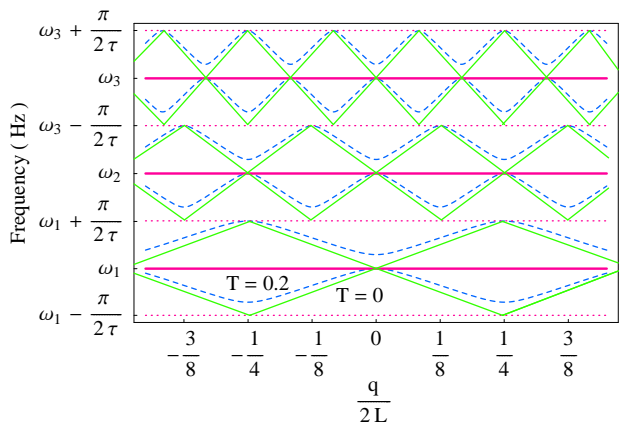}
\caption{\label{fig:transpic2} (Color online). Numerical solution
of Eq.~(\ref{eq:trans}) showing the eigenfrequencies
$\omega_{n}$ of the full 3MC resonator
[Fig.~\ref{fig:transpic}(b)] as a function of middle mirror
position $q$. The solid red seesaw curves are for $T=0$ and the 
dashed blue sinusoidal curves are for $T=0.2$, chosen exaggeratedly 
for visibility.}
\end{figure}

Carrying through the classical calculation the wave vectors $k$
supported by the full resonator appear as solutions to the
transcendental equation \cite{fader1985}
\begin{equation}
 \label{eq:trans}
\cot k(L+q)+\cot k(L-q) = 2\left(\frac{1-T}{T} \right)^{1/2}.
\end{equation}
The solutions of Eq.~(\ref{eq:trans}) imply that as a result of
the coupling between the two sub-cavities of the resonator each 
pair of initially two-fold degenerate modes of frequency $\omega_{n}$
splits into a pair of non-degenerate modes, see
Fig.~\ref{fig:transpic2},
\begin{eqnarray}
 \label{eq:evenodd}
&&\omega_{n,e}(q)\simeq\omega_{n} \nonumber
\\ &+&\frac{1}{\tau}\left[\sin^{-1}\left(\sqrt{1-T}
\cos 2k_{n}q\right)- \sin^{-1}\left(\sqrt{1-T}\right)\right],\nonumber \\
&&\omega_{n,o}(q)\simeq\omega_{n}+
               \frac{\pi}{\tau}\\
               &-&\frac{1}{\tau}\left[\sin^{-1}\left(\sqrt{1-T}
\cos 2k_{n}q\right)+\sin^{-1}\left(\sqrt{1-T}\right)\right]
,\nonumber
\end{eqnarray}
where
\begin{equation}
\label{eq:roundtrip}
\tau=2L/c
\end{equation}
is the round trip time for each sub-cavity, assumed to be
approximately the same for both sides of the resonator for $L\gg
\lambda_n$ and $q \ll \lambda_n$. In Eq.~(\ref{eq:evenodd})
$\omega_{n,e}$ corresponds to a mode with an even number of half
wavelengths in the full resonator, while the mode at frequency
$\omega_{n,o}$ has an additional half-wavelength, hence a slightly
higher frequency. It corresponds to a field maximum at the center
of the resonator, and turns into a `cosine' mode in the limit $T
\rightarrow 1$, while the even mode of frequency $\omega_{n,e}$
turns into a `sine' mode in that limit. The electromagnetic fields
corresponding to the frequencies in Eq.~(\ref{eq:evenodd}) can be
found in Refs.~\cite{chow1986,chow2004}. In these references it is
noted that due to the presence of a `dielectric bump' at the middle
mirror the fields have a discontinuity in their derivative at that
position.

\subsection{Quantization}
\label{subsec:quant}

Sections III-VI concentrate on an analysis restricted to the modes
$\omega_{n,e}$ and $\omega_{n,o}$ about a specific $\omega_{n}$
(Table \ref{tab:mirrorparam}). We quantize these two modes under
the assumption that the oscillation frequency $\omega_M$ of the
middle mirror is sufficiently small that $\tau \ll 1/\omega_M$, so
that the electromagnetic field frequencies follow adiabatically
the mirror motion, and $\omega_{n,e}(q)$ and $\omega_{n,o}(q)$ are
simply parameterized by the mirror position $q$. The Hamiltonian
of the coupled field-mirror system is then
\begin{equation}
\label{eq:hfull} H = \hbar \omega_{\rm e}(q) a^{\dagger} a+ \hbar
\omega_{\rm o}(q) b^{\dagger} b +\frac{p^{2}}{2 m}+\frac{1}{2}m
\omega_M^{2}(q-q_{\rm 0})^{2}
\end{equation}
where we have dropped the subscript $n$ for clarity, $a$ and $b$
are bosonic field operators for the modes of instantaneous
frequencies $\omega_e$ and $\omega_o$ satisfying the commutation
relations
\begin{equation}
 \label{eq:commutation1}
[a,a^{\dagger}]=1, \hspace{0.2in} [b,b^{\dagger}]=1,
\end{equation}
$p$ and $q$ are the momentum and position operators of the moving
mirror, with
\begin{equation}
\label{eq:commutation2} [q,p]=i\hbar,
\end{equation}
and $q_0$ is its rest position in the absence of radiation. The
radiation pressure that couples the mirror motion to the resonator
field is implicitly contained in the position dependence of
$\omega_e$ and $\omega_o$, see Eqs.~(\ref{eq:evenodd}), as we will
see shortly when considering various limits of the Hamiltonian
[Eq.~(\ref{eq:hfull})].

\section{Perfectly reflecting middle mirror}
\label{sec:PRL} For a perfectly reflecting middle mirror, $T=0$,
the even and odd mode frequencies of Eqs.~(\ref{eq:evenodd})
reduce to the eigenfrequencies $\omega_{\rm l,r}$ of the left and
right sub-cavities of the 3MC
\begin{eqnarray}
 \label{eq:wnoT}
\omega_{e} =\omega_{l}&\sim& \omega_{n}\left(1-q/L\right),\nonumber \\
\omega_{o} =\omega_{r}&\sim& \omega_{n}\left(1+q/L\right),
\end{eqnarray}
respectively. These frequencies are shown as the solid red see-saw
lines in Fig.~\ref{fig:transpic2}. For $q_{0}=0$ the Hamiltonian
(\ref{eq:hfull}) can then readily be reexpressed as
\begin{equation}
\label{eq:HamnoT} H = \hbar \omega_{n}(a^{\dagger} a+ b^{\dagger}
b) +\frac{p^{2}}{2 m}+\frac{1}{2}m \omega_M^{2}q^{2} -\hbar \xi
(a^{\dagger} a- b^{\dagger} b)q,
\end{equation}
where
\begin{equation}
 \label{eq:optomechanicalp}
\xi=\omega_{n}/L
\end{equation}
is the optomechanical coupling parameter and $a$ and $b$ are
annihilation operators for the optical modes in the left and right
sub-cavities (Fig.1(c) in \cite{mishpm1}). In this form, the
Hamiltonian (\ref{eq:HamnoT}) shows explicitly the effect of
radiation pressure on the mirror motion. It is the form used in
particular to discuss mirror cooling in Ref.~ \cite{mishpm1}.

We note that the Hamiltonian (\ref{eq:HamnoT}) also holds for
$q_{0} \neq 0$ after a trivial change of coordinate $q \rightarrow
q-q_{0}$. Physically, this indicates that displacing the rest
position of the moving mirror from the center of the resonator
causes no qualitative change in its dynamics. In particular the
radiation pressure term remains linear in the mirror position
(again under the assumption that $q \ll \lambda_{n}$), and the
cooling and trapping behavior is essentially the same as discussed in
Ref.~\cite{mishpm1}. The situation is significantly different for
the case $T\neq 0$, as we now discuss \footnote{We also note that
the experimental arrangement corresponding to this model is that
of the 3MC irradiated from both sides \cite{mishpm1}, see
Fig.\ref{fig:transpic}(b). In this case both end-mirrors have a
small transmissivity $T_{\rm end}$ (Table
\ref{tab:mirrorparam}).}.

\section{$T \neq 0$, linear coupling}
\label{sec:Tnotzero}

As illustrated in Fig.~\ref{fig:transpic2}, the coupling between
the two sub-cavities resulting from the finite transmission of the
moving mirror leads to the appearance of a series of avoided
crossings between $\omega_e(q)$ and $\omega_o(q)$ near those
points where either $\omega_n$ is doubly degenerate for $T=0$, or
two frequencies $\omega_n(q)$ and $\omega_{n'}(q')$ become
degenerate. The slopes of the solid see-saw lines in
Fig.~\ref{fig:transpic2} are given by $\pm \omega_n/L=\pm n\pi
c/L^2$ and are therefore $n$-dependent, hence the anti-crossing
points are not equidistant. For large enough $n$, though, $n
\simeq n+1$ and the avoided crossings occur for mirror separations
$q \simeq \ell {\bar \lambda}/4$ from $q=0$, where $\ell$ is an
integer and ${\bar \lambda}$ is some typical wavelength about
$\lambda_n$.

To lowest order, the dependence of $\omega_e(q)$ and $\omega_o(q)$
on $q$ is linear away from the anti-crossings, but quadratic in
their vicinity. Hence we expect the radiation pressure
contribution to the Hamiltonian (\ref{eq:hfull}) to be likewise
linear and quadratic respectively, in these two cases.

Consider first the linear case where $T \neq 0$ and the rest
position $q_{0}$ of the moving mirror is away from any
anti-crossing point. For small enough mirror displacements, $q \ll
{\bar \lambda}$, we expand $\omega_e(q)$ and $\omega_o(q)$ about
$q_0$ to find
\begin{eqnarray}
 \label{eq:weoT}
\omega_{\rm e} \sim \omega_{n}-\delta_e-\xi_L(q-q_{0}),\nonumber \\
\omega_{\rm o} \sim \omega_{n}+\delta_o+\xi_L(q-q_{0}),
\end{eqnarray}
where
\begin{eqnarray}
\label{eq:Asxis}
\delta_e&=&\frac{1}{\tau}\left[\sin^{-1}\left(\sqrt{1-T}\right)-\sin^{-1}\left(\sqrt{1-T}
\cos 2k_{n}q_{0}\right)\right],\nonumber \\
\delta_o &=& \frac{\pi}{\tau}-\\
& &
\frac{1}{\tau}\left[\sin^{-1}\left(\sqrt{1-T}\right)+\sin^{-1}\left(\sqrt{1-T}
\cos 2k_{n}q_{0}\right)\right],\nonumber \\
\end{eqnarray}
and
\begin{equation}
\xi_L=\frac{\sin2k_{n}q_{0}}{\sqrt{(1-T)^{-1}-\cos^{2}2k_{n}q_{0}}}\xi.\\
\end{equation}
is a generalized linear optomechanical coupling parameter. It is
easy to verify that $|\xi_L| \rightarrow\xi$ for $T=0$. When $T
\neq 0$, $\xi_{L}=0$ for $q_{0}=j \lambda_n/4$, where $j$ is an
integer. This has important consequences that we discuss later on.
A plot of $\xi_L$ is given in appendix \ref{sec:appc},
Fig.~\ref{fig:transpic5}.

With Eqs.~(\ref{eq:weoT}) the Hamiltonian (\ref{eq:hfull}) becomes
\begin{eqnarray}
 \label{eq:HamT}
H&=&\hbar (\omega_{n}-\delta_e)a^{\rm \dagger}a+\hbar (\omega_{n}+\delta_o)b^{\rm \dagger}b \\
&+&\frac{p^{2}}{2 m}+\frac{1}{2}m \omega_M^{2}(q-q_{\rm 0})^{2}
-\hbar \xi_L (a^{\dagger} a- b^{\dagger} b)(q-q_{\rm 0}) \nonumber
\end{eqnarray}
or, with $q-q_0 \rightarrow q$,
\begin{eqnarray}
 \label{eq:HamTsimpler}
H&=&\hbar (\omega_{n}-\delta_e)a^{\rm \dagger}a+\hbar (\omega_{n}+\delta_o)b^{\rm \dagger}b \nonumber \\
&+&\frac{p^{2}}{2 m}+\frac{1}{2}m \omega_M^{2}q^2 -\hbar \xi_L
(a^{\dagger} a- b^{\dagger} b)q.
\end{eqnarray}
We note that this Hamiltonian is \textit{not} equivalent to
setting $q_{0}=0$, since Eq.~(\ref{eq:Asxis}) would then imply
that $\xi_L=0$, and the radiation-mirror coupling would vanish.

Comparing Eqs.~(\ref{eq:HamnoT}) and (\ref{eq:HamTsimpler}) shows
that in the linear coupling regime, the finite mirror transmission
results in the frequencies of the two modes $\omega_{\rm e,o}$
being shifted by $-\delta_e$ and $\delta_o$, respectively, and the
optomechanical constant being redefined as $\xi \rightarrow
\xi_L$. However, since the coupling of the radiation with the
mirror remains linear there is no significant qualitative
difference between the cooling and trapping mechanisms in the
two cases.

Appendix \ref{sec:appc} shows that a simple transformation can put
the dynamical equations for the Hamiltonian
Eq.~(\ref{eq:HamTsimpler}) in a form identical to those for
Eq.~(\ref{eq:HamnoT}), with $\xi$ replaced by $\xi_L$ and
$\delta_{e,o}$ absorbed as detuning shifts. In that same appendix
we show that for an appropriate placement $q_0$ of the mirror we
can obtain $\xi_L \sim \xi$. Hence it should be possible to trap
and cool the partially transparent moving mirror to its quantum
mechanical ground state with essentially the same parameters as
the perfectly reflecting mirror. In Table~\ref{tab:mirrorparam}
we present such a set of parameters. Another set of parameters
has been suggested in Ref.~\cite{thompson2007}.

\section{$T\neq 0$, quadratic coupling}
\label{sec:extrememirr}

\begin{figure}
\includegraphics[width=0.49 \textwidth]{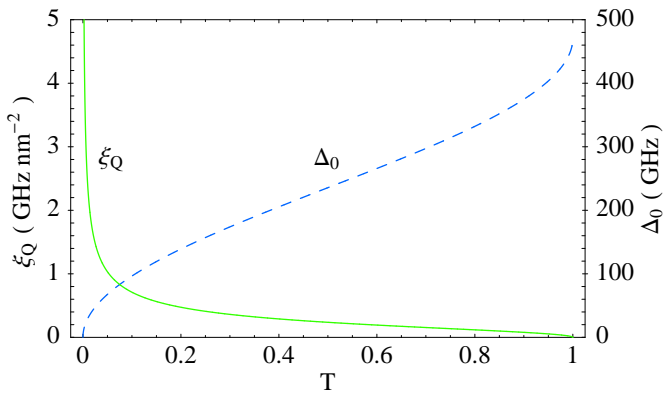}
\caption{\label{fig:transpic3}(Color online). Detuning $\Delta_o$
(blue, dotted line) [Eq.(\ref{eq:del0})] and quadratic optomechanical
coupling constant $\xi_Q$ (red, solid line) [Eq.(\ref{eq:xiq})]
as functions of the middle mirror transmissivity $T$.
The parameter values used to generate these plots are
provided in Table \ref{tab:mirrorparam}.}
\end{figure}

We now turn to the situation where the middle mirror is placed
at a position $q_0=j\lambda_{n}/4$ ($j$ integer). In that case,
expanding Eqs.~(\ref{eq:evenodd}) to lowest order about $q_0$
gives
\begin{eqnarray}
 \label{eq:weoTqex}
\omega_{e} &\sim& \omega_{n}-\xi_Q(q-q_{0})^{2},\nonumber\\
\omega_{o} &\sim& \omega_{n}+\Delta_o+\xi_Q(q-q_{0})^{2},
\end{eqnarray}
where the detuning
\begin{equation}
\label{eq:del0}
\Delta_o=\frac{2}{\tau}
\cos^{-1}(1-T)^{1/2}
\end{equation}
and the quadratic optomechanical coupling constant is
\begin{equation}
\label{eq:xiq}
\xi_Q=\frac{\tau\xi^{2}}{2}\left(\frac{1-T}{T}\right)^{1/2}.\\
\end{equation}
The detuning $\Delta_o$ and $\xi_Q$ are plotted in
Fig.~\ref{fig:transpic3} as functions of $T$.
The Hamiltonian (\ref{eq:hfull}) now becomes
\begin{eqnarray}
 \label{eq:HamTqex}
H&=&\hbar \omega_{n}a^{\rm \dagger}a+\hbar (\omega_{n}+\Delta_o)b^{\rm \dagger}b \\
&+&\frac{p^2}{2 m}+\frac{1}{2}m \omega_M^2(q-q_0)^2 -\hbar \xi_Q
(a^{\dagger} a- b^{\dagger} b)(q-q_0)^2.\nonumber
\end{eqnarray}
Since $\Delta_o$ and $\xi_Q$ are independent of $q_0$ we can
rescale that Hamiltonian by the transformation $q-q_{0}\rightarrow
q$ without affecting any of the physics. This is equivalent to
setting $q_{0}=0$ and yields
\begin{eqnarray}
 \label{eq:HamTqexsimpler}
H&=&\hbar \omega_{n}a^{\rm \dagger}a+\hbar (\omega_{n}+\Delta_o)b^{\rm \dagger}b \nonumber \\
&+&\frac{p^{2}}{2 m}+\frac{1}{2}m \omega_M^{2}q^{2} -\hbar \xi_Q
(a^{\dagger} a- b^{\dagger} b)q^{2}.
\end{eqnarray}
As expected from our previous discussion, the mirror-radiation
coupling is now \textit{quadratic} in the mirror coordinate, in
contrast to Eqs.(\ref{eq:HamnoT}) and (\ref{eq:HamTsimpler}),
where it is \textit{linear} \cite{thompson2007}. This coupling is
purely dispersive and leads to qualitatively different radiation
effects. We show below that such a coupling implies in particular
the ability to trap but not cool the moving mirror.

\section{Effective frequency and damping}
\label{eff-freq}
\subsection{Quantum Langevin Equations} \label{subsec:QLE}
We consider for concreteness a simple implementation of the 3MC
trapping and cooling scheme where the system is driven by a
narrow-band laser field of frequency $\omega_L$ impinging on the
resonator from the left, the right end-mirror being assumed to be
perfectly reflecting. As previously discussed \cite{corbitt2007},
two lasers of different frequencies have to be used in practice to
control the moving mirror. Except for the fact that the powers and
frequencies of these two fields must be chosen self-consistently
in order to ensure the dynamic stability of the system, one of
them essentially affects solely the spring frequency and the other
only the spring damping, so we consider them separately in the
following.

At this point we introduce an additional simplification by noting
that for the value of $T$ considered here we have $\Delta_{o}
\gg \gamma$, (Table \ref{tab:mirrorparam}), that is, the frequency
separation of the two modes is much larger than the cavity
linewidth. In that case, and provided that the laser linewidth is
comparable to or less than $\gamma$, it is sufficient to consider
a single-mode treatment that involves only the resonator mode
closest to $\omega_L$.

By inspection of the last term in Eq.~(\ref{eq:HamTqexsimpler}) we
expect that the `$a$' mode will cause anti-trapping since it is
associated with a negative `spring constant' $-\hbar \xi_{Q}
a^{\rm \dagger} a$ while the mode `$b$' should lead to mirror
trapping. Tuning the laser to the frequency of the latter mode
yields then the approximate single-mode Hamiltonian
\begin{equation}
 \label{eq:HamTsinglemode}
H \simeq\hbar (\omega_{n}+\Delta_o)b^{\rm \dagger}b
+\frac{p^{2}}{2 m}+\frac{1}{2}m \omega_M^{2}q^{2} +\hbar \xi_Q
b^{\dagger} b q^{2}.
\end{equation}
which we analyze below.

The fluctuations of the electromagnetic vacuum couple into the
resonator through the partially transmitting input mirror, which
also leads to the damping of the intracavity field. Further the
Brownian noise associated with the coupling of the oscillating
mirror to its thermal environment must be accounted for in a
realistic treatment of the mirror dynamics. We describe the effect
of these sources of noise and dissipation within the input-output
formalism of quantum optics \cite{gardinerbook}. For the
Hamiltonian (\ref{eq:HamTsinglemode}) this yields in a standard
fashion the nonlinear quantum Langevin equations
\begin{eqnarray}
\label{eq:QLE2}
\dot{b}&=& -\left[i(\delta+\xi_Q q^{2})+\frac{\gamma}{2}\right]b +\sqrt{\gamma} b^{\rm in},\nonumber \\
\dot{q}&=& \frac{p}{m},\nonumber \\
\dot{p}&=& -\left(2\hbar \xi_Q b^{\dagger}b+m\omega_M^{2}\right)q -\frac{D_M}{m}p
          +\epsilon^{\rm in},\\
\end{eqnarray}
where the detuning is given by
\begin{equation}
\label{eq:detuning} \delta = \omega_{n}+\Delta_{o}-\omega_{L},
\end{equation}
and
\begin{equation}
\label{eq:decayratedef} \gamma=\frac{cT_{\rm end}}{2L}
\end{equation}
is the decay rate through the input mirror of transmissivity
$T_{\rm end}$ (Table \ref{tab:mirrorparam}).

In Eq.~(\ref{eq:QLE2}) the noise operator $b^{\rm in}$ describes
the field pumping the cavity mode. It is characterized by the
semiclassical mean value
\begin{equation}
\label{eq:mv}
\langle b^{\rm in}(t) \rangle =b_{s}^{\rm in},
\end{equation}
and Markovian fluctuations
\begin{equation}
 \label{eq:fl}
\langle \delta b^{\rm in}(t) \delta b^{\rm in,\dagger}(t') \rangle=\delta (t-t').
\end{equation}
The Brownian noise operator $\epsilon^{\rm in}$ describes the
heating of the mirror by its thermal environment. It is
characterized by a zero mean value, and fluctuations at
temperature $T_{e}$ correlated as \cite{gardinerbook}
\begin{eqnarray}
\label{eq:coth} &&\langle \delta \epsilon^{\rm
in}(t)\hspace{0.02in}\delta \epsilon^{\rm in}(t') \rangle =
\nonumber \\
&& D_M\int_{-\infty}^{\infty}
\frac{d\omega}{2\pi}e^{-i\omega(t-t')}\hbar \omega \left[1+ \coth
\left(\frac{\hbar \omega}{2k_{\rm B}T_{\rm e}}\right)\right].
\end{eqnarray}
For the parameters of our model, $T_{\rm e} \gg \hbar \omega_{\rm
eff}/k_{\rm B}$, and therefore the high-temperature limit of
Eq.~(\ref{eq:coth}),
\begin{equation}
\label{eq:coth2} \langle \delta \epsilon^{\rm
in}(t)\hspace{0.02in}\delta \epsilon^{\rm in}(t') \rangle= 2
D_M k_{\rm B}T_{\rm e}\, \delta (t-t')
\end{equation}
is applicable.

\subsection{Steady state}
\label{subsec:sse}
Appendix~\ref{sec:appc} shows that for any value of $\xi_Q$,
$q_{s}=0$ is the only real steady-state solution for the mirror
displacement. In contrast to standard configurations bistability
does not occur because we have chosen a trapping mode for the
mirror. The steady-state of the mirror-cavity system is given by
\begin{eqnarray}
\label{eq:ssvalues}
q_{s}&=&0, \,\,\,\,p_{s}=0,\nonumber \\
b_{s}&=&\left[\frac{\gamma}{\delta^{2}+(\gamma/2)^{2}}\right]^{1/2}f_{\rm
s}^{\rm in}, \nonumber \\
\end{eqnarray}
where $f_{s}^{\rm in}=|b_{s}^{\rm in}|$ is the amplitude of the
laser field pumping the cavity. The phase of this field can be
chosen without loss of generality such that $b_{s}$ is real. The
steady-state intracavity field mode amplitude $b_{s}$ in
Eq.~(\ref{eq:ssvalues}) is independent of $\xi_Q$, a consequence
of the fact that $q_{s}=0$.

\subsection{Fluctuations}
\label{subsec:fluc} To account for the effect of the classical and
quantum fluctuations we decompose each operator in
Eq.~(\ref{eq:QLE2}) as the sum of its steady-state value and a
small fluctuation, e.g. $b=b_{s}+\delta b$. Substituting these
quantities into Eq.~(\ref{eq:QLE2}), eliminating the steady-state
contribution and linearizing the resulting equations for the
fluctuations we have
\begin{equation}
\label{eq:linear}
{\dot u}(t)= Mu(t)+n(t).
\end{equation}
Here the vectors of the input noise and fluctuations are respectively given by
\begin{eqnarray}
\label{eq:un}
u(t)&=&(\delta X_{b},\delta Y_{b},\delta q,\delta p),\nonumber \\
n(t)&=&(\sqrt{\gamma}X_{b}^{\rm in},\sqrt{\gamma}Y_{b}^{\rm in},0,\delta \epsilon^{\rm in}),
\end{eqnarray}
and we have symmetrized the fluctuation operators as $\delta X_{b}
=(\delta b+\delta b^{\dagger})/\sqrt{2}$, $\delta Y_{b} =(\delta
b-\delta b^{\dagger})/i\sqrt{2}$, etc. The matrix $M$ is given
explicitly by
\begin{equation}
\label{eq:RHmatrix} M =\\
\begin{pmatrix}
- \gamma/2  &  \delta           &   0                   &   0\\
 -\delta   & -\gamma/2          &  0                    &   0\\
0          &   0                & 0                     &    1/m\\
0 &     0  & -\left(2\hbar \xi_Q b_{s}^{2}+m \omega_M^{2}\right) &-D_M/m\\
\end{pmatrix}.
\end{equation}
The steady-state solutions (\ref{eq:ssvalues}) are dynamically
stable if none of the eigenvalues of the matrix $M$ has a positive
real part. This condition can be quantified in terms of the
Routh-Hurwitz criterion \cite{dejesus1987}, which yields
inequalities too involved to be presented here. However, we will
work at $\delta=0$ in the following and it is quite easy to show
analytically that for this detuning and any value of the other
parameters of Eq.(\ref{eq:RHmatrix}) there is no dynamical
instability in the 3MC.

\subsection{Effective frequency and damping}
\label{subsec:effdf}
In order to determine the effective frequency
$\omega_{\rm eff}$ and damping $D_{\rm eff}$ of the mirror in
the regime of quadratic coupling, Eq.~(\ref{eq:HamTsinglemode}),
we solve the linearized quantum Langevin equations for the
fluctuations in the mirror position,
\begin{equation}
\label{eq:delq} \delta q(\omega) =  \chi (\omega) \,\delta
F_{T}(\omega)
\end{equation}
where $\chi(\omega)$ is the mechanical susceptibility of the
mirror and $\delta F_{T}$, which describes the fluctuations in the
total force on the mirror, consists of a radiation vacuum and a
Brownian motion component. The susceptibility has the form of a
Lorentzian \cite{cohadon1999}
\begin{equation}
\label{eq:chi} \chi^{-1} (\omega) =m(\omega_{\rm
eff}^{2}-\omega^{2})-iD_{\rm eff}\omega,
\end{equation}
from which we can extract the effective oscillation frequency
$\omega_{\rm eff}$ and damping constant $D_{\rm eff}$ of the
mirror as
\begin{eqnarray}
    \label{eq:eff}
\omega_{\rm eff}^{2} &=&\omega_M^{2}+\frac{(2\xi_Q\gamma
P_{\rm in}/ m\omega_{n} )}{\delta^{2}+(\gamma/2)^{2}},\nonumber \\
D_{\rm eff}&=&D_M,
\end{eqnarray}
where $ P_{\rm in}=\hbar\omega_{n}|f_{s}^{\rm in}|^{2}$ is the
input power of the incident laser. Strikingly, $D_{\rm eff}$ is
unchanged from its intrinsic value, signifying the absence of the
usual back-action effects responsible for mirror damping.
Specifically we see that if we retain only terms linear in the
fluctuations the quadratic optomechanical coupling of
Eq.~(\ref{eq:HamTqexsimpler}) does not affect the damping of the
mirror. This is because the radiation-mirror coupling is purely
dispersive, as pointed out in Ref.~\cite{thompson2007}.

The modification of the mirror frequency in Eq.~(\ref{eq:eff}) is
a result of the dependence of the cavity mode frequencies on the
position $q$ of the moving mirror. Radiation pressure trapping
translates into an increase of the effective mirror frequency from
its `bare' value $\omega_M$. It follows from Eq.~(\ref{eq:eff})
that the mirror can be optically trapped by selectively exciting
the mode $\omega_{\rm o}$, and that the trapping effect is
strongest on resonance $\delta=0$, with the resulting trapping
frequency
\begin{equation}
 \label{eq:wmax}
\omega_{\rm max}^{2}=\omega_M^{2}+\frac{4\xi_{Q}P_{\rm in}}{m
\omega_{n}\gamma}.
\end{equation}
For this configuration there is neither (static) bistability nor
dynamical instability, hence a high laser power can be used to
achieve tight mirror traps, limited only by the effects of mirror
heating.

In Eq.~(\ref{eq:eff}) the effective frequency depends on the
mirror transmissivity through the coupling parameter $\xi_{Q}$
[Eq.~(\ref{eq:xiq})] which can be large for small values of $T$ as
evident from Fig.~\ref{fig:transpic3}. For the parameters of this
paper the effective frequency $\omega_{\rm eff}$ turns out to be
much larger than the bare mechanical frequency $\omega_M$ (see
below) and it is comparable to the effective frequency achieved in
the 3MC with a perfectly reflecting middle mirror (Eq.~(7) in
Ref.~\cite{mishpm1}). However, the trapping light does not
introduce any anti-damping in the present case, in contrast to
both the $T=0$ situation and the linear coupling regime described
by the Hamiltonian (\ref{eq:HamTsimpler}).

For completeness we mention that an equivalent single-mode
treatment of the coupling of the incident laser into the even mode
of frequency $\omega_{e}$ leads to anti-trapping and
instabilities, both static as well as dynamic. We do not consider
this regime further in this paper.

\section{Bichromatic trapping and cooling}
\label{sec:propbi}

\begin{figure}
\includegraphics[width=0.49 \textwidth]{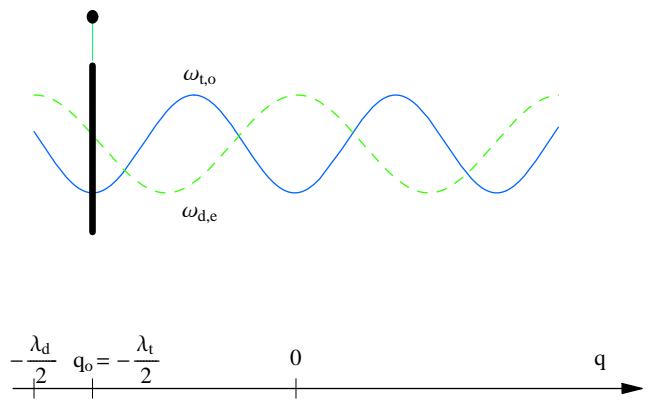}
\caption{\label{fig:transpic4}(Color online) Schematic of the
proposed two-color cooling and trapping scheme. The region close to the
center of the cavity $(q=0)$ is shown. The sinusoidal curves
correspond to the frequencies of two resonator modes as a function
of the middle mirror displacement $q$ from the origin. An odd mode
of frequency $\omega_{t,o}$ (solid, blue line) excited by a laser
of wavelength $\lambda_{t}$ and an even mode of frequency
$\omega_{d,e}$ (dotted, red line) excited by a second laser of
wavelength $\lambda_{d}$ are shown. The equilibrium position $q_0$
of the mirror is chosen so as to coincide with a minimum of the
$\omega_{t,o}$ mode and to be slightly to the right of a maximum
of the $\omega_{d,e}$ mode. Red-detuning the wavelength
$\lambda_{d}$ of the second laser damps the mirror motion via a
linear optomechanical coupling and tuning the wavelength
$\lambda_{t}$ of the trapping field to resonance traps the mirror
via a quadratic optomechanical coupling. }
\end{figure}

Summarizing our results so far, we have shown in
Section~\ref{sec:Tnotzero} and Appendix~\ref{sec:appb} that in the
regime of linear optomechanical coupling an even mode can be used
to achieve passive cooling, while Section ~\ref{subsec:effdf}
demonstrates the possibility of achieving a large effective
frequency $\omega_{\rm eff}$ without introducing any anti-damping
in the regime of quadratic optomechanical coupling. These results
suggest the use of two incident lasers at wavelengths
$\lambda_{d}$ and $\lambda_{t}$ that drive the mirror in the
linear and quadratic coupling regimes, respectively, to damp and
trap its motion, see Fig.~\ref{fig:transpic4}. These wavelengths
are chosen such that the corresponding resonant cavity mode
numbers $n_{d,t} \gg 1$ so that $n_{d,t} \sim n_{d, t} +1$. We
also assume that each incident laser is effectively coupled to
only one cavity mode so that the single mode treatments of
sections~\ref{subsec:effdf} and \ref{sec:Tnotzero} and
appendix~\ref{sec:appb} are valid. In order for the two lasers to
act essentially independently on the moving mirror we also require that
they couple into resonator modes whose frequency separation is
much larger than $\gamma$.

The dependence of the frequencies $\omega_{t,o}$ and
$\omega_{d,e}$ of the two relevant cavity modes on the mirror
position $q$ is as illustrated in Fig.~\ref{fig:transpic2}, but
with the mutual shift in position shown schematically in
Fig.~\ref{fig:transpic4}. The moving mirror is at a position such
that $\omega_{t,o}$ has an extremum (the first minimum to the left
of $q=0$, at $q_0= -\lambda_t/2$ in Fig.~\ref{fig:transpic4}.). The
laser excitation of that mode results in mirror trapping with no
anti-damping, as we have seen.

The first maximum of $\omega_{d,e}$ is at $-\lambda_{d}/2$ from
the cavity center, hence the mirror is at a distance
$(\lambda_{d}-\lambda_{t})/2$ to the right of that maximum, see
Fig.~\ref{fig:transpic4}. Appendix \ref{sec:appb} shows that
regular passive cooling can be implemented by red-detuning the
$\lambda_{d}$ radiation from $\omega_{d,e}$, and that in order to
optimize radiation effects for $T \sim 10^{-4}$ the mirror should
be displaced by an amount of the order of $\lambda_{d}/10$ to the
\textit{right} of the maximum of $\omega_{d,e}$, (Fig.\ref{fig:transpic5}).
This implies 
\begin{equation}
 \label{eq:l1l2}
(\lambda_{d}-\lambda_{t})/2 =\lambda_{d}/10,
\end{equation}
which gives $\lambda_{t}=0.8 \, \lambda_{d}$ for the example of
Fig.~\ref{fig:transpic4}. In an actual experiment, an appropriate
$q_{0}$ can be found empirically given two available laser
wavelengths.

This `hybrid' configuration enables a trap stiffness unrestricted
by considerations of anti-damping. In other words the trapping
light at $\lambda_{t}$ does not destabilize or raise the noise
temperature of the middle mirror at all. This technique is
therefore superior to the standard trapping and cooling scheme
based only on a radiation-mirror coupling linear in the mirror
coordinate, such as Eqs.~(\ref{eq:HamnoT}) or
(\ref{eq:HamTsimpler}). Both the effective mirror damping 
(due to absence of anti-damping) as well as the mirror trapping
(due to absence of instabilities) can be stronger in the hybrid 
case, and the achievable final degree of vibrational excitation 
[Eq.~(\ref{eq:nosc})] is therefore lower.

For example, if anti-damping is absent in the case of purely
linear coupling treated in appendix~\ref{sec:appb}, the total
damping increases by a factor of $\sim 100$, implying a mirror
temperature lower by the same factor. The trapping required to
reach the ground state now can be achieved by 8mW of laser light
on resonance.

\section{Conclusion}
\label{sec:conclusion} The radiative trapping and cooling of a
totally or partially reflecting mirror in an optical cavity has
been considered theoretically. Our main conclusion is that
allowing the middle mirror to be quite transmissive does not
greatly affect the ability of radiation to cool the mirror down to
its quantum mechanical ground state. In fact the parameters
required to accomplish ground state occupation for the transparent
mirror are virtually the same as for a perfectly reflecting
mirror. For this reason the advantages of the 3MC over the 2MC
mentioned in Ref.~\cite{mishpm1} are retained even if the middle
mirror is allowed to be a little transparent.

We have shown that the nature of the mirror-radiation interaction
can be changed from dissipative to dispersive depending on the
position of the middle mirror with respect to the end mirrors of
the cavity, in agreement with the analysis of
Ref.~\cite{thompson2007}, and also that in the dispersive regime
strong optical trapping of the mirror is possible without any
anti-damping.

Combining the various regimes of optomechanical coupling that we
have identified, we have also proposed a novel two-color mirror
trapping and cooling scheme based on positioning the mirror so as
to simultaneously couple it dissipatively with one cavity mode and
dispersively with a second mode. In contrast to all the
configurations implemented or discussed in the literature so far,
trapping in this configuration does not cause anti-damping or
instabilities of either the static or dynamic kind. This improves
the damping effect of radiation while allowing for tighter mirror
traps to be established using higher laser power. This allows to
reach lower mirror temperatures and eases the route to the
occupation of the quantum mechanical ground state of the moving
mirror.

\begin{table*}
\begin{center}
\renewcommand{\arraystretch}{1.7}
\begin{tabular*}{0.597\textwidth}{|c|c|c|c|c|}
\hline \textbf{No.} & \textbf{Parameter} & \textbf{Description} & \textbf{Value} & \textbf{Units} \\
\hline
\hline
1.                  & $L$                & sub-cavity length                                  & 5 &  mm         \\
\hline
2.                  & $\lambda$          & laser wavelength                                   & 514   &  $\mu$m         \\
\hline
3.                  & $n$                & mode number                                        & 10$^{4}$   &     -      \\
\hline
4.                  & $\omega_{n}$       & cavity resonance frequency                         & 2$\pi$10$^{15}$   &  Hz        \\
\hline
5.                  & $\xi$             & optomechanical coupling parameter                    & 100 & MHz nm$^{-1}$   \\
\hline
6.                  & $\xi_L$             & linear optomechanical coupling                    & 100 & MHz nm$^{-1}$   \\
\hline
7.                  & $T$                 & middle mirror transmissivity                       & 10$^{-4}$ &        -            \\
\hline
8.                  & $\Delta_o$          &  mode frequency shift                              & 1 & GHz          \\
\hline
9.                  & $\xi_Q$              & quadratic optomechanical coupling                & 100 &  MHz nm$^{-2}$    \\
\hline
10.                  & $m$                 & middle mirror mass                                 & 1 &  $\mu$g                \\
\hline
11.                 & $\omega_M$          & middle mirror resonance frequency                  & 2$\pi$2.5   &  kHz        \\
\hline
12.                 & $D_M$              & middle mirror damping constant                     & 0.02 &  $\mu$g Hz     \\
\hline
13.                 & $T_{e}$            & middle mirror initial temperature              & 300 &  K                         \\
\hline
14.                 & $T_{\rm end}$          & end mirror transmissivity                          & 10$^{-5}$ &  -                  \\
\hline
15.                 & $\gamma$           & cavity linewidth                    & 2$\pi$5 &  MHz                  \\
\hline
\end{tabular*}
\end{center}
\caption{\label{tab:mirrorparam} Definitions and approximate
values of some of the parameters used in the text.}
\end{table*}

\begin{acknowledgments}
This work is supported in part by the US Office of Naval Research,
by the National Science Foundation and by the US Army Research
Office. We thank O. Dutta, Dr. C. Maes and Prof. H. Ritsch for
useful discussions.
\end{acknowledgments}
\appendix
\section{Effective temperature and quanta}
\label{sec:appa} This Appendix derives Eqs.~(\ref{eq:lowT}) and
(\ref{eq:nosc}) and discusses their limit of validity. We mention
here a correction to Eq.(1) of Ref.~\cite{mishpm1} which presented
an incorrect scaling of the mirror quanta with effective
frequency, and thus underestimated the degree of excitation of the
mirror. This correction however does not bring any qualitative
change to our previous results \cite{mishpm1}. That same
correction has recently been realized by other authors
\cite{karrai2004,marquardt2007}.

The starting point of our derivation is Eq.~(\ref{eq:delq}).
For the parameters of the
model discussed in Sections~\ref{sec:PRL}, ~\ref{sec:Tnotzero}
and ~\ref{sec:extrememirr},
the fluctuations in the total force are due mainly to thermal
noise, so that $\delta F_{\rm T}(\omega) \sim \delta \epsilon^{\rm
in}(\omega)$. In the case of Section~\ref{sec:extrememirr} this is
exactly true i.e. $\delta F_{\rm T}(\omega) \equiv \delta
\epsilon^{\rm in} (\omega)$ since fluctuations in the radiation
field do not couple to the mirror motion in the framework of
linear response theory. In either case the two-time correlation
function of the force fluctuations is therefore that of the
thermal component,
\begin{equation}
\label{eq:forcecorr}
\langle \delta F_{\rm T}(t) \,\delta F_{\rm T}(t') \rangle =
\langle \delta \epsilon^{\rm in}(t) \, \delta \epsilon^{\rm in}(t')
\rangle= N \delta(t-t'),
\end{equation}
where $N=2 D_{M} k_{\rm B} T_{\rm e}$ according to
Eq.~(\ref{eq:coth2}).

Fourier transforming (FT) both sides of Eq.~(\ref{eq:forcecorr})
using the symmetric FT
\begin{equation}
 \label{eq:FT}
\delta F_{\rm T}(\omega) = \frac{1}{\sqrt{2\pi}}
\int_{-\infty}^{\infty} dt \,e^{i\omega t} \,\delta F_{\rm T}(t)
\end{equation}
gives the frequency-domain correlation function
\begin{equation}
\label{eq:forcecorrom}
\langle \delta F_{\rm T}(\omega) \,\delta F_{\rm T}(\omega') \rangle =
N \delta(\omega+\omega'),
\end{equation}
which in turn allows us to express the correlation function for
the linear displacement, see Eq.~(\ref{eq:delq}), as
\begin{equation}
\label{eq:lincorr}
\langle \delta q(\omega)\, \delta q(\omega') \rangle =
N \chi(\omega)\,\chi(\omega')\,\delta(\omega+\omega').
\end{equation}
Inverse Fourier transforming both sides of Eq.~(\ref{eq:lincorr})
we get
\begin{equation}
 \label{eq:lintcorr}
\langle \delta q(t) \,\delta q(t') \rangle =
\frac{N}{2\pi} \int_{-\infty}^{\infty} d\omega \,e^{-i\omega(t-t')}
\,|\chi(\omega)|^{2},
\end{equation}
since $\chi(-\omega)=\chi^{*}(\omega)$ from Eqs.~(\ref{eq:chi}),
(\ref{eq:eff}) and (\ref{eq:effapp}). Setting $t=t'$ in
Eq.~(\ref{eq:lintcorr}) we get
\begin{equation}
\label{eq:deltaqsq}
\langle \delta q^{2}(t) \rangle =
\frac{N}{2\pi} \int_{-\infty}^{\infty} d\omega \,|\chi(\omega)|^{2}.
\end{equation}
We now use the equipartition theorem to link the average displacement
squared to $T_{\rm eff}$, the effective temperature of the vibrating
mirror
\begin{equation}
\label{eq:equi} \frac{k_{\rm B}T_{\rm eff}}{2} = \frac{m
\omega_{\rm eff}^{2}\langle \delta q^{2}(t)\rangle}{2}.
\end{equation}
Note that the equipartition theorem is expressed in terms of the
effective frequency of the mirror. Combining
Eqs.(\ref{eq:deltaqsq}) and (\ref{eq:equi}) we find
\begin{equation}
\label{eq:Teff} T_{\rm eff}=T_{\rm e} \left(\frac{m
\omega_{\rm eff}^{2}D_{M}}{\pi}\right) \int_{-\infty}^{\infty}
d\omega \,|\chi(\omega)|^{2}.
\end{equation}
This allows us to determine the mean number of quanta of vibration
of the moving mirror as
\begin{equation}
\label{eq:quantaapp} n_M=\frac{k_{\rm B} T_{\rm eff}}{\hbar
\omega_{\rm eff}}= \frac{k_{\rm B} T_{\rm e}}{\hbar}
\left(\frac{m \omega_{\rm eff}D_{M}}{\pi}\right)
\int_{-\infty}^{\infty} d\omega \,|\chi(\omega)|^{2}
\end{equation}
where
\begin{equation}
\label{eq:suscep}
\chi^{-1}(\omega) = m[\omega_{\rm eff}^{2}(\omega)-\omega^{2}]-
iD_{\rm eff}(\omega)\omega.
\end{equation}
The exact form of the effective frequency and damping depend on
the position and transmission of the middle mirror. They are given
by either Eq.~(\ref{eq:eff}) or Eq.~(\ref{eq:effapp}). In the case
of Eq.~(\ref{eq:eff}) the effective quantities do not depend on
$\omega$. In the case of Eq.~(\ref{eq:effapp}) we expand them in a
Taylor series around $\omega=\omega_M$ and keep only the leading
terms in the respective expansions (see below for justification),
\begin{eqnarray}
 \label{eq:Taylor1}
\omega_{\rm eff}^{2}(\omega) &\sim& \omega_{\rm eff}^{2}(\omega_M)\equiv \omega_{\rm eff}^{2},\nonumber \\
D_{\rm eff}(\omega) &\sim& D_{\rm eff}(\omega_M) \equiv
D_{\rm eff}.
\end{eqnarray}
With Eq.~(\ref{eq:Taylor1}) we find analytically
\begin{equation}
\label{eq:suscan} \int_{-\infty}^{\infty} d\omega
\,|\chi(\omega)|^{2}=\frac{\pi}{m(\omega_{\rm eff})^{2}
D_{\rm eff}}
\end{equation}
Equations (\ref{eq:Teff}) and (\ref{eq:quantaapp}) then imply
\begin{equation}
\label{eq:lowT2} T_{\rm eff}=\left(\frac{D_M}{D_{\rm eff}}\right)T_{\rm e}
\end{equation}
and
\begin{equation}
\label{eq:nosc2} n_{M}=\frac{k_{\rm B} T_{\rm eff}}{\hbar
\omega_{\rm eff}}= \frac{k_{\rm B} T_{\rm e}}{\hbar \omega_{\rm eff}}\left(\frac{D_M}{D_{\rm eff}}\right),
\end{equation}
respectively. These are precisely Eqs.~(\ref{eq:lowT}) and
(\ref{eq:nosc}). Evidently Eqs.~(\ref{eq:suscan}),
(\ref{eq:lowT2}) and (\ref{eq:nosc2}) are exact when the effective
frequency and damping follow from Eq.~(\ref{eq:eff}). (We will say
no more about this case.). These results are however
approximate when the effective frequency and damping follow from
Eq.~(\ref{eq:effapp}).

To ensure that the approximation stated in Eq.~(\ref{eq:Taylor1})
is accurate, we used the full functional forms of $\omega_{\rm
eff}(\omega)$ and $D_{\rm eff}(\omega)$ from
Eq.~(\ref{eq:effapp}), and performed the integral in
Eq.~(\ref{eq:suscan}) numerically. For the parameters used in this
paper this yields numerical values indistinguishable from the
approximate analytical expressions, i.e. we found the same
effective temperature and quanta for the mirror.

The condition that needs to be satisfied for the approximation in
Eq.~(\ref{eq:Taylor1}) to be valid can be found by inspecting the
forms of the functions $\omega_{\rm eff}(\omega)$ and $D_{\rm
eff}(\omega)$ in Eq.~(\ref{eq:effapp}). Expanding analytically
these expressions in Taylor series about $\omega_M$ and defining
\begin{equation}
 \label{eq:V}
V=\frac{4\xi \gamma P_{\rm in}}{mL},
\end{equation}
we find
\begin{equation}
 \label{eq:taylorweff}
\omega_{\rm eff}^{2}(\omega)=\omega_{\rm
eff}^{2}(\omega_M)+d(\omega_M) (\omega-\omega_M)+\mathcal{O}
[(\omega-\omega_M)^{2}],
\end{equation}
where \begin{widetext}
\begin{eqnarray}
 \label{eq:t1}
\omega_{\rm
eff}^{2}(\omega_M)&=&\omega_M^{2}-\frac{16V\delta(-4\omega_M^{2}+\gamma^{2}+4\delta^{2})}
{(\gamma^{2}+4\delta^{2})\left[16\omega_M^{4}+8\omega_M^{2}(\gamma^{2}-4\delta^{2})+(\gamma^{2}+4\delta^{2})^{2}\right]}
\nonumber \\
          &\simeq&  -\frac{16V\delta(-4\omega_M^{2}+\gamma^{2}+4\delta^{2})}
{(\gamma^{2}+4\delta^{2})\left[16\omega_M^{4}+8\omega_M^{2}(\gamma^{2}-4\delta^{2})+(\gamma^{2}+4\delta^{2})^{2}\right]}
\end{eqnarray}
is the first term in the expansion and $\omega_M^{2}$ can be
neglected since we are in a regime where the optical contribution
to the stiffness is typically much larger than the intrinsic
mechanical contribution (this may be translated into an appropriate
condition on $V$ [Eq.~(\ref{eq:V})]). The coefficient of the
second term in Eq.~(\ref{eq:taylorweff}) is
\begin{equation}
 \label{eq:d}
 d(\omega_M)=\frac{-128V\omega_M\delta\left[16\omega_M^{4}-3
\gamma^{4}-8\gamma^{2}\delta^{2}+16\delta^{4}-8\omega_M^{2}(\gamma^{2}+4\delta^{2})\right]
}{(\gamma^{2}+4\delta^{2})
\left[16\omega_M^{4}+8\omega_M^{2}(\gamma^{2}-4\delta^{2})+(\gamma^{2}+4\delta^{2})^{2}\right]^{2}}.
\end{equation}
The contribution of the second term becomes comparable to that of
the first in Eq.~(\ref{eq:taylorweff}) at the critical frequency
\begin{equation}
 \label{eq:wcrit}
\omega_{\rm
c}=\omega_M-\frac{(4\omega_M^{2}-\gamma^{2}-4\delta^{2})
\left[16\omega_M^{4}+8\omega_M^{2}(\gamma^{2}-4\delta^{2})+(\gamma^{2}+4\delta^{2})^{2}\right]}
{8\omega_M\left[16\omega_M^{4}-3
\gamma^{4}-8\gamma^{2}\delta^{2}+16\delta^{4}-8\omega_M^{2}
(\gamma^{2}+4\delta^{2})\right]}.
\end{equation}
\end{widetext}
Using the hierarchy $\delta \geq \gamma/2 \gg \omega_M$ applicable to
this article, we approximate Eq.~(\ref{eq:wcrit}) and get
\begin{equation}
 \label{eq:wcrittrap}
\omega_{\rm c} \simeq
\omega_M\left[1+\frac{1}{2}\left(\frac{\delta}{\omega_M}\right)^{2}\right].
\end{equation}
The detuning is usually a few cavity linewidths for trapping, e.g.
$\delta = -2.5 \gamma$ in this work. Thus $\delta / \omega_M \sim
10^{3}$ and therefore $\omega_{\rm c} \sim 10^{6}\omega_M$. Also,
for our parameters $\omega_{\rm eff}(\omega_M)\lesssim
10^{3}\omega_M$. This implies that not only is the critical
frequency much larger than the intrinsic mechanical frequency, it
is also much larger than the optically-induced mirror frequency,
an important observation. A similar result can be obtained for
$D_{\rm eff}$, in which case $\delta = \gamma/2.$ The general
conclusion is that for our parameters the higher order
frequency-dependent terms in the expansions of $\omega_{\rm
eff}(\omega)$ and $D_{\rm eff}(\omega)$ become important at
frequencies much higher than $\omega_{\rm eff}(\omega_M)$.

Now for $\omega_{\rm eff}(\omega) \simeq \omega_{\rm
eff}(\omega_M)$ [from Eq.~(\ref{eq:Taylor1})] the integrand in
Eq.~(\ref{eq:Teff}) is a Lorentzian peaked at $\omega_{\rm
eff}(\omega_M)$, with a symmetric peak at -$\omega_{\rm
eff}(\omega_M)$. The regime $|\omega| \geq \omega_{c}$ then
corresponds to the far-out wings of the Lorentzian, since
$\omega_{c} \gg \omega_{\rm eff}(\omega_M)$. In this regime the
contribution of the higher-order terms to the spectrum is highly
suppressed, resulting in virtually no change in the area
underneath the spectrum and no change in the scalings in
Eqs.~(\ref{eq:lowT2}, \ref{eq:nosc2}). Thus the condition under
which Eqs.~(\ref{eq:lowT}) and (\ref{eq:nosc}) hold is
\begin{equation}
\label{eq:condition} \omega_{\rm eff}(\omega_M) \ll \omega_{c},
\end{equation}
which is well satisfied in our case.

\section{Linear optomechanical coupling}

\label{sec:appb} This appendix discusses the two-mode situation
described by the linear coupling Hamiltonian
(\ref{eq:HamTsimpler}), and relates it to the case of a 3MC with
perfectly reflecting middle mirror of Ref. \cite{mishpm1}. We
consider specifically the situation where the frequencies of the
two resonator modes under consideration are widely separated and
are driven by two independent incident lasers of frequencies
$\omega_{La}$ and $\omega_{Lb}$.

The quantum Langevin equations for the Hamiltonian of
Eq.~(\ref{eq:HamTsimpler}) can be written as
\begin{eqnarray}
\label{eq:QLET}
\dot{a}&=&-\left[i(\delta-\xi_{L} q)+\frac{\gamma}{2}\right]a+\sqrt{\gamma} a^{\rm in},\nonumber \\
\dot{b}&=&-\left[i(\delta+\xi_{L} q)+\frac{\gamma}{2}\right]b+\sqrt{\gamma} b^{\rm in},\nonumber \\
\dot{q}&=& p/m,\nonumber \\
\dot{p}&=& -m\omega_M^{2}q+\hbar \xi_{L} (a^{\dagger}a-b^{\dagger}
b) -\frac{D_M}{m}p+\epsilon^{\rm in},
\end{eqnarray}
where $\gamma$ describes the decay rates of the left and right
sub-cavities,, taken to be equal for simplicity, and the
frequencies of the two laser have been chosen such that
\begin{equation}
\label{eq:detuningT}
\delta=\omega_{n}-\delta_{e}-\omega_{La}=\omega_{n}+\delta_{o}-\omega_{Lb}.
\end{equation}
Equation~(\ref{eq:QLET}) is exactly the same as Eq.~(3) of
Ref.~\cite{mishpm1} with the change of notation $\xi \rightarrow
\xi_{L}$, hence it yields the same radiation effects.

\begin{figure}
\includegraphics[width=0.45 \textwidth]{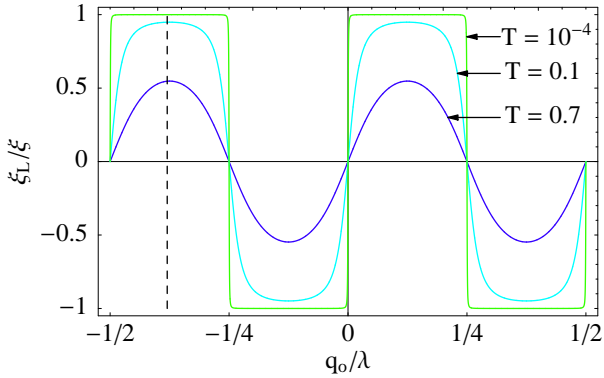}
\caption{\label{fig:transpic5}(Color online) Linear optomechanical
coupling parameter $\xi_{L}$ as a function of the middle mirror
placement $q_{0}$ for various values of the mirror transmission
$T$. The dotted line indicates the position of the mirror at
$q_{0}=-\lambda/2+ \lambda/10$, allowing $\xi_{L}$ to approach
closely the maximum value $\xi$ corresponding to a perfectly
reflecting mirror.}
\end{figure}

The effective parameters for the middle mirror can therefore be
determined from Eq.~(7) of Ref.~\cite{mishpm1} which we reproduce
below
\begin{widetext}
\begin{eqnarray} \label{eq:effapp} \omega_{\rm
eff}^{2}&=&\omega_M^{2}-\frac{4\xi \gamma P_{\rm in}}{mL}
\frac{\delta}{\delta^{2}+\frac{\gamma^{2}}{4}}
\frac{(\frac{\gamma}{2})^{2}-(\omega^{2}-\delta^{2})}
{\left[(\frac{\gamma}{2})^{2}+(\omega-\delta)^{2}\right]
\left[(\frac{\gamma}{2})^{2}+(\omega+\delta)^{2}\right]},\nonumber \\
D_{\rm eff}&=&D_M+\frac{4\xi \gamma P_{\rm in}}{L}
\frac{\delta}{\delta^{2}+\frac{\gamma^{2}}{4}}
\frac{\gamma}{\left[(\frac{\gamma}{2})^{2}+(\omega-\delta)^{2}\right]
\left[(\frac{\gamma}{2})^{2}+(\omega+\delta)^{2}\right]}.
\end{eqnarray}
\end{widetext}
Fig.~\ref{fig:transpic5} shows that for an appropriate middle
mirror position we can have $\xi_{L} \sim \xi$. Even for a mirror
transmission as large as $T=0.7$ \cite{thompson2007} $\xi_{L} \sim
\xi/2$ is possible, indicating that the cooling of the moving
mirror to its ground state of vibration is possible both for
weakly transparent as well as for perfectly reflecting mirrors
using essentially the same parameters. It also follows that the
advantages of the 3MC over the 2MC pointed out in
Ref.~\cite{mishpm1} are retained even in the case where the middle
mirror is partially transparent.

Table~\ref{tab:mirrorparam} lists the relevant numerical values
considered in this paper. Using 5mW of trapping light at a
detuning $\delta_{\rm t}=-2.5\gamma$ and $10\mu$W of cooling
light at a detuning $\delta_{\rm c}=0.5\gamma$ we find
$\omega_{\rm eff} \sim 300\omega_{M}$,
$D_{\rm eff} \sim 10^{6}D_{M}$ and $T_{\rm eff} \sim 200\mu$K.
Here we have assumed a mechanical quality factor of $10^{6}$,
an optical finesse of $10^{5}$ and an ambient temperature of
300K. From Eq.~(\ref{eq:nosc}), these values imply $n_{M}<1$.

We note that in Eq.~(3) of Ref.~\cite{mishpm1} $a$ and $b$ are the
annihilation operators of the modes in the sub-cavities, while in
Eq.~(\ref{eq:QLET}) they correspond to modes of the full
resonator; in the case of a finite transmission, both modes need
to be pumped to obtain a behavior analogous to that of the 3MC
with the perfectly reflective middle mirrors.

We finally remark that if we do not pump the odd mode [i.e. set
$b\equiv 0$ in Eq.~(\ref{eq:HamTsimpler})], we obtain the 2MC
Hamiltonian
\begin{equation}
\label{eq:Honlyeven} H = \hbar
\omega_{c}a^{\dagger}a+\frac{p^{2}}{2m} +\frac{1}{2}m
\omega_M^{2}Q^{2}-\hbar\xi'a^{\dagger}aQ,
\end{equation}
and hence can trap or cool with a single mode. In our proposal for
a hybrid design for cooling and trapping in
Section~\ref{sec:propbi}, we red-detune around the even mode to
achieve passive cooling.

\section{Steady state solutions and bistability}
\label{sec:appc} This appendix considers the bistability of the
steady-state solutions in the case of a partially transparent
mirror placed in such a way that the resonator frequency is at a
minimum, see Eq.~(\ref{eq:ssvalues}).

The equations obtained by setting the time derivatives in
Eq.~(\ref{eq:QLE2}) equal to zero are
\begin{eqnarray}
\label{eq:ssapp}
b_{\rm s}&=&\frac{\sqrt{\gamma}b_{\rm s}^{\rm in}}{i(\delta+\hbar \xi_{Q}q_{s}^{2})+\frac{\gamma}{2}},\nonumber \\
p_{\rm s}&=&0, \nonumber \\
0&=&-\left(\hbar \xi_{Q}|b_{\rm s}|^{2}+m
\omega_M^{2}\right)q_{\rm s},
\end{eqnarray}
where we have solved for $p_{\rm s}$. The value of $b_{\rm s}$
can be obtained by using the first equation in the last
and solving for $q_{\rm s}$.

From the last equation we see that $q_{\rm s}=0$ is the only real
solution for the mirror position. This is because the factor in
parentheses is the sum of two positive non-zero terms, and can
never equal zero for real $q_{s}$. Hence there is no bistability
for the single-mode configuration of Section~\ref{subsec:sse}.

\bibliography{Mirrors}
\end{document}